\documentclass[aps,twocolumn,showpacs,amsmath,amsfonts,amssymb]{revtex4}
\usepackage{graphicx}
\usepackage{epsf}

\begin{document}
\title{Mesoscopic superposition and sub-Planck scale structure in molecular wave packets}
\author{Suranjana Ghosh$^{\mathrm{1}}$\footnote{e-mail:
sanjana@prl.ernet.in}, Aravind
Chiruvelli$^{\mathrm{2}}$\footnote{e-mail:
Aravind.Chiruvell001@students.umb.edu}, J.
Banerji$^{\mathrm{1}}$\footnote{e-mail: jay@prl.ernet.in} and P.
K. Panigrahi$^{\mathrm{1}}$\footnote{e-mail:
prasanta@prl.ernet.in}}

\affiliation{$^{\mathrm{1}}$Physical Research Laboratory,
Navrangpura, Ahmedabad-380 009, India\\$^{\mathrm{2}}$University
of Massachusetts at Boston, Boston, MA 02125-3393, USA}

\begin{abstract}
We demonstrate the possibility of realizing sub-Planck scale
structures in the mesoscopic superposition of molecular wave
packets involving vibrational levels. The time evolution of the
wave packet, taken here as the SU(2) coherent state of the Morse
potential describing hydrogen iodide molecule, produces cat-like
states, responsible for the above phenomenon. We investigate the
phase space dynamics of the coherent state through the Wigner
function approach and identify the interference phenomena behind
the sub-Planck scale structures. The optimal parameter ranges are
specified for observing these features.

\end{abstract}
 \maketitle
%
%
%
%
Mesoscopic superposition of coherent states and their
generalizations, such as cat-like states, have attracted
considerable attention in the recent literature
\cite{schleich,haroche,tara}, since they show a host of
non-classical behaviors. In a remarkable paper, Zurek \cite{zurek}
demonstrated that appropriate superposition of some of these
states with a
 classical action $A$ can lead to sub-Planck scale structures in
phase space. These sub-Planck scale structures in the phase space
are characterized by an area $\hbar^{2}/A$.  Apart from their
counter intuitive nature and theoretical significance, the above
scale has been shown to control the effectiveness of decoherence,
a subject of tremendous current interest in the area of quantum
computation and information. Zurek's realization made use of
dynamical systems which exhibit chaotic behavior in the classical
domain. Recently a cavity QED realization involving the mesoscopic
superposition of the compass states have been given \cite{pathak}.
In principle, one could also use superpositions of cat-like states
arising in quantum optical systems with large Kerr non-linearity
\cite{tara}.

In this paper, we demonstrate the possibility of realizing
sub-Planck scale structures in the mesoscopic superposition of
molecular wave packets,  which involves vibrational levels. The
time evolution of an initial wave packet, taken here as the SU(2)
coherent state (CS) of the Morse potential produces cat-like states.
These arise due to the quadratic dependence of the energy on the
vibrational quantum number. The superposition of these states is
responsible for the above phenomena. We study the spatio-temporal
structure of these states, paying special attention to the
 fractional revival, which gives rise to four
coherent states required for the observation of the sub-Planck
structure. This structure can be clearly explained through the
interference phenomena in phase space. For this, we investigate
the phase space dynamics of the coherent state through the Wigner
function approach and identify the optimal parameter ranges for a
clear observation of these features.

Morse potential is well-known to capture the vibrational dynamics
of a number of diatomic molecules
 \cite{vetchin1,vetchin2,morse,dahl,wolf}. It is worth mentioning that
the phenomena of revival and fractional revival
\cite{averbukh,bluhm,robinett} have been experimentally observed
in wave packets involving vibrational levels \cite{stolow}.
Creation of the wave packets and observation of their dynamics are
carried out through pump-probe method \cite{wilson}. The control
and analysis of molecular dynamics is achieved through ultrashort
femto-second laser pulses \cite{garraway}. Fractional revival can
be probed by random-phase fluorescence interferometry
\cite{warmuth}. Recently, cat-like states, arising in the temporal
evolution of the Morse system, have been proposed for use in the
quantum logic operations \cite{shapiro1}.

Morse potential describing the vibrational motion of a diatomic
molecule has the form
\begin{equation}
V(x)=D(e^{-2\beta x}-2e^{-\beta x})
\end{equation}
 where
$x=r/r_{0}-1,\;r_{0}$ is the equilibrium value of the
inter-nuclear distance $r$, $D$ is the dissociation energy and
$\beta$ is a range parameter. We will be considering HI molecule,
as an example, which has $30$ bound states, with $\beta=2.07932$,
reduced mass $\mu=1819.99\;a.u.$, $r_{0}=3.04159\;a.u.$ and
$D=0.1125\;a.u.$ Defining
\begin{equation}
\lambda=\sqrt{\frac{2\mu D r^{2}_{0}}{\beta^2\hbar^2}}\;\;\mathrm{and}\;\;
s=\sqrt{-\frac{8\mu r^{2}_{0}}{\beta^2\hbar^2}E},
\end{equation}
eigen functions of the Morse potential can be
written as
\begin{equation}\label{eigenstate}
\psi_{n}^{\lambda} (\xi)= N e^{-\xi/2} \xi^{s/2}
L_{n}^{s} (\xi),
\end{equation}
where $\xi=2\lambda e^{-\beta x}; \;\;0<\xi<\infty$, and
$n=0,1,...,[\lambda-1/2]$, with $[\rho]$ denoting the largest
integer smaller than $\rho$, so that the total number of bound
states is $[\lambda-1/2]$. The parameters $\lambda$ and $s$
satisfy the constraint condition $s+2n=2\lambda-1$.

Note that $\lambda$ is potential dependent, $s$ is
related to energy $E$ and, by definition,  $\lambda>0,\;s>0$. In Eq.~(\ref{eigenstate}),
 $L_{n}^{s}(y)$ is the associated Laguerre polynomial and N is the
normalization constant:

\begin{equation}
N=\left[\frac{\beta(2\lambda-2n-1)\Gamma{(n+1)}}{\Gamma{(2\lambda-n)}r_{0}}\right]^{1/2}.
\end{equation}

\begin{figure}
\includegraphics[width=3in]{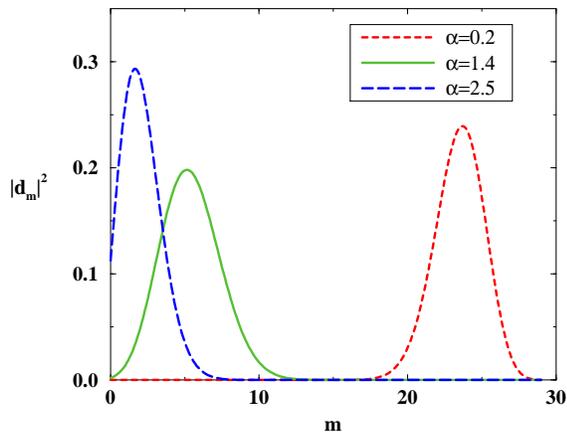}
\caption{(Color online) \label{dm}$|d_{m}|^2$ plotted as a function of $m$ for the Morse potential of HI
molecule for different values of $\alpha$.}
\end{figure}
\ Quite some time back, Nieto and Simmons gave a minimum
uncertainty coherent state for Morse oscillator considering
suitable conjugate variables \cite{nieto}. Later, Benedict and
Moln\'{a}r \cite {benedict} also found the same CS through super
symmetric quantum mechanical method. This was used to describe the
cat states of the NO molecule \cite{molnar}. This CS involves
infinite number of bound states, not belonging to the same
potential \cite{charan}. Morse potential has a finite number of
bound states. Hence it is natural to expect an underlying SU(2)
algebra. Recently, Dong \emph{et al}., \cite{frank} have obtained
the SU(2) generators $\hat{J}_{+}, \hat{J}_{-}$ and $\hat{J}_{0}$
which satisfy the algebra at the level of wave function as
\begin{equation}
\left[\hat{J}_{+},\hat{J}_{-}\right]\psi_{n}^{\lambda}
(\xi)=2\hat{J}_{0}\psi_{n}^{\lambda} (\xi).
\end{equation}

The SU(2) Perelomov coherent state of the Morse system is obtained
by operating the displacement operator $e^{\alpha
\hat{J}_{+}-\alpha^*\hat{J}_{-}}$ on the highest bound state $n'$,
defined by $\hat{J}_{+}\psi_{n'}^{\lambda}(\xi)=0$. Using
disentanglement theorem, the coherent state modulo normalization
becomes
\begin{eqnarray}
\chi(\xi)&=& e^{-\alpha
\hat{J}_{-}} \psi_{n'}^{\lambda}(\xi)\nonumber\\
&&=\left[\psi_{n'}^{\lambda}-\alpha
\sqrt{n'(s+ n'+1)}\;\psi_{n'-1}^{\lambda}+...\right.\nonumber\\
&&+\left.\frac{(-\alpha)^{n'}}{({n'})!}\sqrt{n'!
(s+n'+1)(s+n'+2)}\right.
\nonumber\\
&&\hspace{2cm}\left.\times\sqrt{....(s+2n')}\;\;\psi_{0}^{\lambda}\right],
\end{eqnarray}

As is explicitly seen the above CS involves only the bound states,
which are finite in number. This is due to the fact that the
underlying group here is a compact group \cite{perelomov}. For the
purpose of our analysis, we consider this wave packet. We have
checked that, superposition of Morse eigen states with suitable
Gaussian weight factors, also reproduces the sub-Planck scale
structure.

Simplifying the above expression, we can write it in a compact
form:
\begin{equation}
\chi(\xi)=\sum_{m=0}^{n'}d_{m} \;\psi_{m}^{\lambda}(\xi),
\end{equation}
where
\begin{equation}
d_{m}=\frac{(-\alpha)^{n'-m}}{(n'-m)!}\left[\frac{n'!
\Gamma(2\lambda-m)} {m!\Gamma(2\lambda-n')}\right]^\frac{1}{2}.
\end{equation}

 Fig.~\ref{dm} shows $|d_{m}|^2$ distribution of HI molecule for various values of
 $\alpha$. For smaller values of $\alpha$, $|d_{m}|^2$ is peaked at
 higher values of $m$, where the anharmonicity is larger. The
 corresponding initial CS wave packet is not well localized and has an oscillatory tail.
 With the increase of $\alpha$, $|d_{m}|^2$
 distribution moves towards the lower levels  and the wave packet's
 oscillatory tail gradually disappears. For larger values of
$\alpha$, only the lower levels contribute to form the CS wave
packet, where the effect of anharmonicity is rather small. Hence,
it is clear that the choice of the distribution is quite crucial
in the wave packet localization and its subsequent dynamics.

\begin{figure*}
\centering
\includegraphics[width=6.7in]{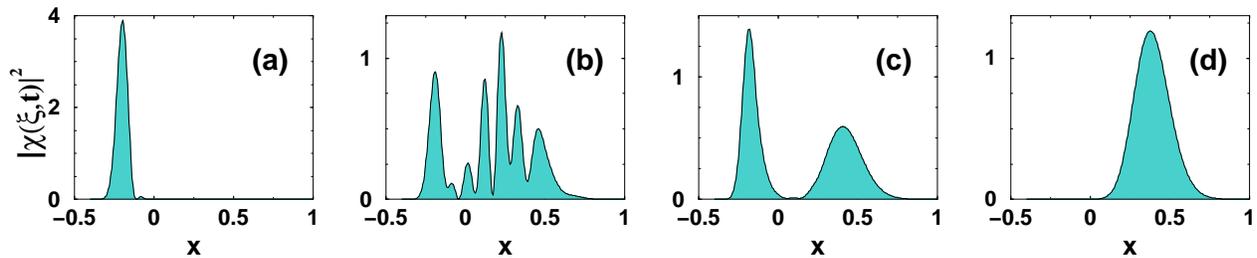}
\caption{(Color online) Wave packet distribution in coordinate space for HI
molecule, where $\alpha=1.4,\; \beta=2.07932$. Plotted here is $\vert \chi(\xi,t)\vert^2$ as a function of $x$ (where $\xi=2\lambda \exp[-\beta x])$ for (a) t=0, (b)
$t=T_{\mathrm{rev}}/8$, (c)
$t=T_{\mathrm{rev}}/4$ and (d)
$t=T_{\mathrm{rev}}/2$.}\label{prob2D}
\end{figure*}

Temporal evolution of CS state wave packet is given by
\begin{equation}\label{cswavepacket}
\chi(\xi,t)=\sum_{m}d_{m}\psi_{m}^{\lambda}(\xi) \exp[-iE_{m}t]
\end{equation}
with $ E_m=-(D/\lambda^{2})(\lambda-m-1/2)^2$. This quadratic
energy spectrum yields classical and the revival times given by
 $T_{\mathrm{cl}}=T_{\mathrm{rev}}/(2 \lambda-1)$ and
$T_{\mathrm{rev}}=2\pi\lambda^{2}/D$ respectively. More
interestingly, when $\mathrm{t}$ takes the values
 $\frac{r}{q}T_{\mathrm{rev}}$, where $r$ and $q$ are mutually prime
integers, the CS wave packet can be written as a sum of
 classical CS wave packets \cite{averbukh}:
\begin{equation}\label{cscl}
\chi(\xi,t)=\sum_{p}^{l-1}a_{p}\;\chi_{\mathrm{cl}}[\xi,(r/q\;
T_{\mathrm{rev}}-p/l\;T_{\mathrm{cl}})],
\end{equation}
 where
 \begin{equation}
\chi_{\mathrm{cl}}(\xi,t)=\sum_{m}d_{m}\psi_{m}^{\lambda}(\xi)
\exp[-2\pi imt/T_{\mathrm{cl}}].
\end{equation}
Amplitudes are determined by
\begin{equation}
a_{p}=\frac{1}{l}\sum_{m}^{l-1}\exp\left[2\pi
i(m^2r/q-mp/l)\right],
\end{equation}
where $l=q/2$ when $q$ is an integer multiple of $4$ and $l=q$, in
all other cases.

Fig.~\ref{prob2D} shows the CS wave packet in the co-ordinate
representation, where the revival behaviors at
$T_{\mathrm{rev}}/4$ and $T_{\mathrm{rev}}/8$ are not transparent.
We will now clarify the phase space picture of the wave packet at
fractional revival times by using the Wigner function approach. We
will also show that the interference phenomenon in phase space
involving the cat-like states gives rise to the sub-Planck scale
structure.

The Wigner function can be written as

\begin{eqnarray}\label{wigner}
W(x,p,t)&=&\frac{r_{0}}{\pi\hbar}\int_{-\infty}^{+\infty}
\bar{\chi}^{*}(x-x',t)\nonumber\\&& \times\bar{\chi}(x+x',t)
e^{-2ipx'/\hbar}dx'\;,
\end{eqnarray}
where $x$ is the scaled co-ordinate and $p$ is the corresponding
scaled momentum and also
 $\bar{\chi}(x)=\chi(\xi)$.

Wigner functions at instances of fractional revival can be
explained by making use of the decomposition of Eq.~(\ref{cscl}).
At $t=T_{\mathrm{rev}}/8$, for example, the CS wave packet splits
into four classical wave packets:

\begin{eqnarray}\label{class}
\chi(\xi,\frac{T_{\mathrm{rev}}}{8}
)&=&\frac{1}{2}[e^{i\pi/4}\chi_{cl}(\xi,\frac{T_{\mathrm{rev}}}{8}
)\nonumber\\&
+&\chi_{cl}(\xi,\frac{T_{\mathrm{rev}}}{8}-\frac{T_{\mathrm{cl}}}{4}
)\nonumber\\&-&e^{i\pi/4}\chi_{cl}(\xi,\frac{T_{\mathrm{rev}}}{8}
-\frac{T_{\mathrm{cl}}}{2})\nonumber\\&+&\chi_{cl}(\xi,\frac{T_{\mathrm{rev}}}{8}
-\frac{3T_{\mathrm{cl}}}{4} )]. \;
\end{eqnarray}

Defining
\begin{equation}
\chi^{\mathrm{(even,odd)}}_{\mathrm{cl}}(\xi,t)=\sum_{{\mathrm{m_{
even,odd}}}}d_{m}\psi_{m}^{\lambda}(\xi) \exp[-2\pi
im\frac{t}{T_{\mathrm{cl}}}]
\end{equation}
expression Eq.~(\ref{class}) can be rewritten in a simpler form:
\begin{equation}\label{classwig}
\chi(\xi,\frac{T_{\mathrm{rev}}}{8})=\chi^{\mathrm{even}}_{\mathrm{cl}}(\xi,\frac{T_{\mathrm{rev}}}{8}-\frac{T_{\mathrm{cl}}}{4})
+e^{i\pi/4}\chi^{\mathrm{odd}}_{\mathrm{cl}}(\xi,\frac{T_{\mathrm{rev}}}{8}).
\end{equation}

\begin{figure*}
\centering
\includegraphics[width=9.0in]{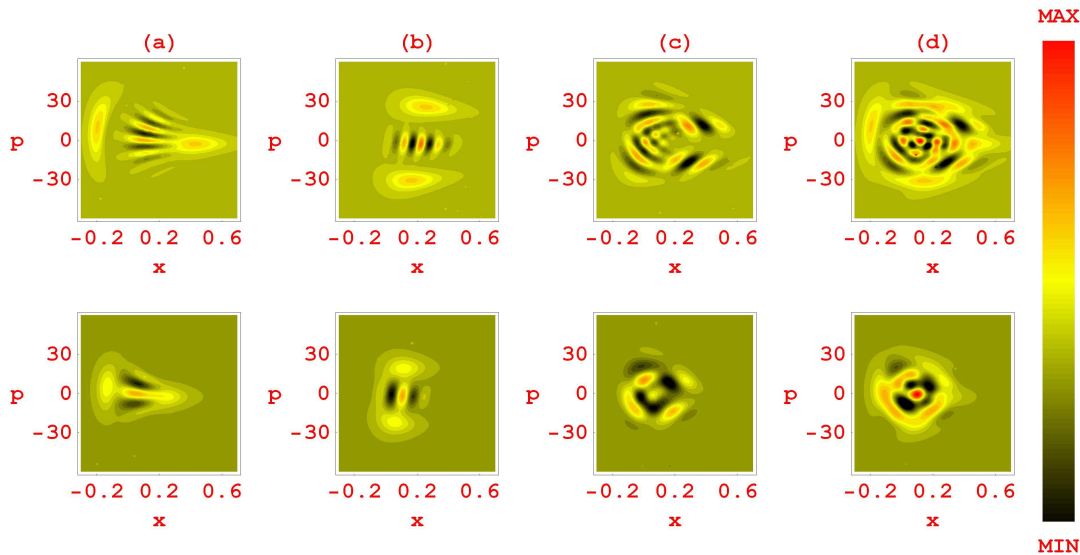}
\caption{(Color online) The Wigner function $W(x,p,t)$ and its
constituent parts at $t=T_{\mathrm{rev}}/8$ as a function of $x$
and $p$ for $\alpha=1.4$ (top row) and $\alpha=2.5$ (bottom row).
Shown here are the contour plots of (a)$W^{(\mathrm{even})}$;
(b)$W^{(\mathrm{odd})}$; (c)$W^{(\mathrm{int})}$ and
(d)$W(x,p,t)$.}\label{contclass}
\end{figure*}

The above expression plays a crucial role in the explanation of the
 phase space behavior at $T_{\mathrm{rev}}/8$. Substituting this in Eq.~(\ref{wigner}), the Wigner function at
$t=T_{\mathrm{rev}}/8$ can be written down as a sum of three
terms: $W(x,p,T_{\mathrm{rev}}/8)=W^{(\mathrm{even})}
+W^{(\mathrm{odd})}+W^{(\mathrm{int})}$, where
$W^{(\mathrm{even})}$ and $W^{(\mathrm{odd})}$ are the Wigner
functions corresponding to the first and second terms on the right
hand side of Eq.~(\ref{classwig}) and $W^{(\mathrm{int})}$ is the
contribution from the interference between these two terms. In
Fig.~\ref{contclass}, we have plotted $W(x,p,T_{\mathrm{rev}}/8)$
and its constituent parts for two values of $\alpha$. Note that
both $W^{(\mathrm{even})}$ and $W^{(\mathrm{odd})}$ are Wigner
functions of cat-like states. Each consists of two distinct peaks
corresponding to two mesoscopic wave packets, and an oscillatory
structure at the middle due to quantum interference between them.
Furthermore, $W^{(\mathrm{even})}$ is along the east-west
direction whereas $W^{(\mathrm{odd})}$ is along the north-south.
This is because the time arguments of
$\chi^{\mathrm{even}}_{\mathrm{cl}}$ and
$\chi^{\mathrm{odd}}_{\mathrm{cl}}$ differ by $T_{\mathrm{cl}}/4$
in Eq.~(\ref{classwig}). The superposition of the interference
regions of $W^{(\mathrm{even})}$ and $W^{(\mathrm{odd})}$ gives
rise to the sub-Planck structure in Fig.~\ref{contclass}(d). It is
worth pointing out that $W^{\mathrm{int}}$, as plotted in
Fig.~\ref{contclass}(c), gives the off diagonal interferences of
compass-like states produced at $T_{\mathrm{rev}}/8$.

 As seen in Fig.~\ref{dm}, for higher values of $\alpha$, the initial wave
packet involves lower vibrational levels for which the turning
points are nearer, resulting in a decrease in the span of the
phase space variables. In this case, the area of overlap between
the two interference structure increases and the number of ripples
become less. As a consequence, the sub-Planck scale structure at
the middle becomes more prominent as shown in the bottom array of
Fig.~\ref{contclass}. The four mini-wave packets, produced at
$T_{\mathrm{rev}}/8$, are not equi-spaced and not of same size.
The asymmetrical nature of the Morse potential is the main reason
behind this. We also analyze numerically the expectation values of
position and momentum at $t=T_{\mathrm{rev}}/8$ for different
values of $\alpha$. The uncertainty product $(\triangle x\triangle
p)$, obtained from this analysis, is $5.5914$ for $\alpha=1.4$ and
$2.56404$ for $\alpha=2.5$ in the unit of $\hbar=1$. The classical
action is defined by $A\approx \triangle x\triangle p$ and the
corresponding dimension of the sub-Planck scale structure is
$a\approx\hbar^{2}/A$ \cite{zurek}, which easily comes out to be
$0.179$ for $\alpha=1.4$ and $0.39$ for $\alpha=2.5$ respectively,
implying the sub-Planck scale structure. Note that for smaller
values of $\alpha$ the area becomes more sub-Planck.

In conclusion, we demonstrate that the interesting sub-Planck
structure in mesoscopic quantum systems can indeed be realized in
the temporal evolution of vibrational wave packets. This is
clearly present, where four wave packets are produced in the
temporal evolution. The coherence parameter $\alpha$ plays a
crucial role in the formation of this structure; one needs the
low-lying states for a clear observation of this structure. It is
worth pointing out that, the vibrational wave packets are prone to
decoherence through coupling to rotational and other vibrational
levels. Recently methods like closed-loop control \cite{brif} has
been devised to minimize the decoherence effect.

\end{document}